\documentclass[a4paper,11pt]{article}
\usepackage{pos}
\usepackage{lineno}
\usepackage{cleveref}
\crefname{section}{Sec.}{Secs.}
\crefname{figure}{Fig.}{Figs.}
\crefname{equation}{Eq.}{Eqs.}
\crefname{table}{Table}{Tables}
\crefname{appendix}{Appendix}{Appendices}

\newcommand{\BackgroundRandomNuENH}{$0.052^{+0.035}_{-0.022}$}
\newcommand{\BackgroundRandomNuMuNH}{$0.34^{+0.15}_{-0.11}$}
\newcommand{\BackgroundRandomNuENC}{$0.014^{+0.024}_{-0.011}$}

\newcommand{\NexpNuE}{$4.08_{-0.94}^{+2.84}$ (flux) $_{-0.25}^{+0.25}$ (cross-section) $_{-0.87}^{+1.24}$ (other syst)}
\newcommand{\NexpNuMu}{$21.15_{-1.90}^{+3.54}$ (flux) $_{-1.27}^{+1.27}$ (cross-section) $_{-4.43}^{+6.47}$ (other syst)}

\newcommand{\NncContami}{$0.1290 _{-0.0197}^{+0.0417}$ (flux) $_{-0.0077}^{+0.0077}$ (cross-section) $_{-0.0684}^{+0.2033}$ (other syst)}

\title{Latest neutrino results from the FASER experiment and their implications for forward hadron production}
\ShortTitle{Latest neutrino results from FASER and their implications for forward hadron production}

\author*[a]{Ken Ohashi}
\affiliation[a]{Albert Einstein Center for Fundamental Physics, Laboratory for High Energy Physics, \\University of Bern, Sidlerstrasse 5, CH-3012 Bern, Switzerland}

\author[b]{Tomohiro Inada}
\affiliation[b]{Kyushu University, Nishi-ku, 819-0395 Fukuoka, Japan}

\author[c,d]{Felix Kling}
\affiliation[c]{Deutsches Elektronen-Synchrotron DESY, Notkestr.~85, 22607 Hamburg, Germany}
\affiliation[d]{Department of Physics and Astronomy, University of California, Irvine, CA 92697-4575, USA}

\author[d]{Max Fieg}

\author[]{the FASER Collaboration}

\emailAdd{ken.ohashi@cern.ch}

\abstract{\ \ \ \ \ 
The muon puzzle --- an excess of muons relative to simulation predictions in ultra-high-energy cosmic-ray air showers ---  has been reported by many experiments. 
This suggests that forward particle production in hadronic interactions is not fully understood. 
Some of the scenarios proposed to resolve this predict reduced production of forward neutral pions and enhanced production of forward kaons (or other particles).
The FASER experiment at the LHC is located 480~m downstream of the ATLAS interaction point and is sensitive to neutrinos and muons, which are the decay products of forward charged pions and kaons. 
In this study, the latest measurements of electron and muon neutrino fluxes are presented using the data corresponding to 9.5~$\mathrm{fb^{-1}}$ and 65.6~$\mathrm{fb^{-1}}$ of proton-proton collisions with $\sqrt{s}=13.6~\mathrm{TeV}$ by the FASER$\nu$ and the FASER electronic detector, respectively. These fluxes are compared with predictions from recent hadronic interaction models, including \texttt{EPOS-LHCr}, \texttt{SIBYLL 2.3e}, and \texttt{QGSJET 3}.
The predictions are generally consistent with the measured fluxes from FASER, although some discrepancies appear in certain energy bins. 
More precise flux measurements with additional data will follow soon, enabling validation of pion, kaon, and charm meson production with finer energy binning, reduced uncertainties, and multi-differential analyses.
}

\ConferenceLogo{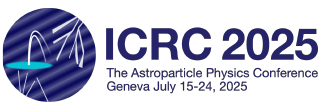}

\FullConference{39th International Cosmic Ray Conference (ICRC2025)\\
 15–24 July 2025\\
Geneva, Switzerland\\}

\begin{document}
\maketitle

\section{Introduction}




Neutrinos at the LHC arise primarily from decays of the lightest hadrons --- namely charged pions, charged and neutral kaons, and charm mesons. The forward production of these particles at LHC energies has not been directly measured. 
Thus, neutrino flux measurements by a forward experiment at the LHC provide a novel probe of forward hadron production, complementing previous measurements of neutral particles by the LHCf experiment~\cite{LHCf:2017fnw,LHCf:2018gbv,Piparo:2023yam}.
The FASER experiment is a far-forward experiment at the ATLAS interaction point and is sensitive to neutrinos from proton-proton collisions, thereby offering unique insights into otherwise inaccessible physics.

The forward production of light hadrons lies outside the domain of perturbative quantum chromodynamics (QCD) and is typically described by phenomenological hadronic interaction models. Prominent and widely used examples include \texttt{EPOS-LHC}~\cite{Pierog:2013ria}, \texttt{SIBYLL~2.3d}~\cite{Riehn:2019jet}, and \texttt{QGSJET~2.04}~\cite{Ostapchenko:2010vb}. 
Earlier this year, updated versions of these generators --- \texttt{EPOS-LHCr}~\cite{Pierog:2023ahq,EPOSICRC2025}, \texttt{SIBYLL~2.3e}, and \texttt{QGSJET~3}~\cite{Ostapchenko:2024myl} ---  were released. 
These models have been tuned and validated using a wide range of data, including results from low-energy experiments, central LHC detectors, and LHCf. FASER's measurements of neutrino fluxes contribute complementary data sets sensitive to the forward production of charged pions, charged and neutral kaons as well as charm hadrons. 

Improving forward particle production models is essential for astroparticle physics, where they are used to describe the production of high-energy particles in extreme astrophysical environments and the interactions of cosmic rays with Earth’s atmosphere. 
In particular, understanding the production of pions and kaons is crucial for addressing the muon puzzle, which refers to the fact that, for many years, experimental measurements of the number of muons in high- and ultra-high-energy cosmic-ray air showers have been in tension with model predictions~\cite{Soldin:2021wyv}. 
One viable class of scenarios proposed as a solution predicts the enhancement of kaons, which can be detected using the neutrino flux measurements from the FASER experiment~\cite{Anchordoqui:2022fpn}.

Recently, the FASER experiment reported measurements of the $\nu_\mu$ and $\bar{\nu}_\mu$ fluxes at the LHC~\cite{FASER:2024ref, FASER:2927629}. 
In addition, the observations of electron and muon neutrinos using the FASER$\nu$ emulsion detector have been reported in Ref.~\cite{FASER:2024hoe} and were recently updated in Ref.~\cite{Ariga:2927714}. 
Notably, electron neutrino flux measurements can provide a unique constraint on kaon and charm hadron production. 
In this study, the $\nu_e + \bar{\nu}_e$ and $\nu_\mu + \bar{\nu}_\mu$ fluxes are calculated using the latest FASER$\nu$ results presented in Moriond 2025~\cite{Ariga:2927714}. The fluxes are presented alongside the previously measured neutrino fluxes. 
Moreover, in order to facilitate a comparison of the combined FASER flux measurements with different hadronic interaction models, the results are presented in a single figure.

\section{FASER detector}

The FASER detector is located 480~m downstream of the ATLAS interaction point at the LHC behind a 100~m of bedrock. 
From upstream to downstream, the FASER detector comprises a FASER$\nu$ front veto scintillator system, the FASER$\nu$ emulsion-tungsten detector, the veto scintillator system, the decay volume, the timing scintillator, three tracking stations, the preshower scintillator system, and the calorimeter~\cite{FASER:2022hcn}.
Three 0.57~T dipole magnets are located in the detector: one at the decay volume and the other two between the tracking stations. The magnets and three tracking stations function as a spectrometer and measure the momentum and charge of charged particles. 
The FASER$\nu$ detector consists of 730 layers of emulsion films interleaved with tungsten plates. 
The 1.1~ton FASER$\nu$ detector is used as a target mass for neutrino analyses.
FASER has published neutrino measurements using two methods. 
The first method uses the electronic detector to select muon appearance events, which are detected by requiring no signal in the FASER$\nu$ front veto scintillator system and a clear muon signal in the spectrometer~\cite{FASER:2023zcr, FASER:2024ref, FASER:2927629}. 
The other method involves reconstructing neutral vertices, using charged particle tracks, found in the emulsion detector, to select neutrino interaction candidates~\cite{FASER:2024hoe, Ariga:2927714}.


\section{Neutrino flux measured by the FASER$\nu$ detector}
\subsection{Dataset and simulation}


In this analysis, the dataset and neutrino candidates reported at Moriond 2025~\cite{Ariga:2927714} are used. A portion of the second FASER$\nu$ detector, which recorded data in 2022 corresponding to an integrated luminosity of $9.5~\mathrm{fb^{-1}}$ of proton-proton collisions with $\sqrt{s}=13.6~\mathrm{TeV}$, is analyzed.
The emulsion films in the detector were extracted and processed at the Emulsion Facility at CERN and scanned in Nagoya University, and the tracks were reconstructed using FASER$\nu$ reconstruction software~\cite{FASER:2025qaf}.
The total target mass of the analyzed volume is 324.1 kg, after data quality requirements, which is 2.45 times larger than that used in the first neutrino result using the FASER$\nu$ detector~\cite{FASER:2024hoe}. 
The analysis area corresponds to the pseudo-rapidity region $\eta \geq 8.76$. 

The expected number of charge-current (CC) neutrino events, $N_{\nu}$, is calculated using neutrino flux predictions, cross-section models, and estimated detection efficiencies as follows:
\begin{equation}
    N_{\nu} = \frac{L\,\rho\,l}{m_{\mathrm{nucleon}}}
    \int \mathrm{d}^2 x\,\mathrm{d}E \bigl(\phi^{\nu}(E, \vec{x})\,\sigma^{\nu}(E, \vec{x})\,\epsilon^{\nu}(E, \vec{x}) 
    + \phi^{\bar{\nu}}(E, \vec{x})\,\sigma^{\bar{\nu}}(E, \vec{x})\,\epsilon^{\bar{\nu}}(E, \vec{x})\bigr),
    \label{eq:number_neutrino}
\end{equation}
where $L$, $\rho$, and $l$ denote the integrated luminosity, target tungsten density, and target thickness, respectively.
Here, $\phi^\nu$ and $\phi^{\bar{\nu}}$ are the neutrino and antineutrino fluxes, respectively, and $\sigma$ and $\epsilon$ denote the corresponding cross-sections and detection efficiencies, respectively.
Integration is performed over the neutrino energy (E) and detector surface area ($\vec{x}$) (note this corresponds to the neutrino rapidity). 
The neutrino fluxes and cross-sections are predicted following the standard methodology used by the FASER Collaboration described in Ref.~\cite{FASER:2024ykc}. The applied event selections are summarized in the next section. The detection efficiency after applying selections was evaluated in Ref.~\cite{FASER:2024hoe}. Table~\ref{tab:summay_signal_expectation} summarizes the number of expected $\nu_e+\bar\nu_e$ and $\nu_\mu + \bar\nu_\mu$ events passing their respective selections. 
%

\begin{table}
\centering
\caption{Number of CC events expected to pass the respective selection cuts. The uncertainties of expectations from flux, cross-section, and others are also shown. 
}  
    \begin{tabular}{l|cc}
    \hline\hline
        Interaction type & Value \\
        \hline
        CC $\nu_e$  & \NexpNuE \\
        CC $\nu_\mu$ & \NexpNuMu \\
    \hline\hline
    \end{tabular}
    \label{tab:summay_signal_expectation}
\end{table}

\subsection{Neutrino event search}


Neutrino event searches are performed using the method described in Ref.~\cite{FASER:2024hoe}.
First, neutral vertices, formed from a converging pattern of tracks with an impact parameter of $5\,\mathrm{\mu m}$ without any upstream charged-particle tracks, are found using charged-particle tracks passing through at least three films. The vertices are further selected by requiring at least three tracks with $\tan \theta > 0.1$, where $\theta$ is the polar angle measured from the beam, and starting within three films downstream of a vertex. A selection on $\tan \theta$ is applied to suppress the neutral hadron backgrounds.
From the selected vertices, the following selections are applied to find $\nu_e$ or $\nu_\mu$ CC candidates. An EM shower with energy $E>200~\mathrm{GeV}$ and $\tan \theta > 0.005$, starting within two films downstream from the vertex, is required for $\nu_e$ CC candidates. A muon penetrating 100 films without significant secondary interactions and with momentum $p_\mu > 200~\mathrm{GeV}$ and $\tan \theta > 0.005$ is required for $\nu_\mu$ CC candidates. For both cases, $\Delta \phi > \frac{\pi}{2}$, where $\Delta \phi$ is the relative azimuthal angle between the lepton and other vertex-associated tracks in the $x$–$y$ plane, is required to suppress the neutral current interaction backgrounds.
Five $\nu_e+\bar{\nu}_e$ and 19 $\nu_\mu+\bar{\nu}_\mu$ candidate events are identified~\cite{Ariga:2927714}.

\subsection{Neutrino flux measurements}



The neutrino interaction rate is given by the product of the cross-section and neutrino flux in Eq.~\ref{eq:number_neutrino}. Using the predicted Standard Model cross-section, the neutrino flux from the measured interaction rate can be inferred. 
In this study, the $\nu_e + \bar{\nu}_e$ and $\nu_\mu + \bar{\nu}_\mu$ interaction rates are interpreted by comparing the number of observed events with expectations.

The total number of observed events, $N$, is given by the sum of the signal (neutrino CC) and background contributions, namely neutrino Neutral-Current (NC) events and neutral-hadron interactions.
The expected neutral hadron background, $N^{\rm NH}$, is estimated to be \BackgroundRandomNuENH \, for electron neutrino candidates and \BackgroundRandomNuMuNH \, for muon neutrino candidates, obtained by scaling the target mass of the background estimates from the previous analysis~\cite{FASER:2024hoe}. 
The contribution of each neutral hadron is summarized in Table II of Ref.~\cite{FASER:2024hoe}. 
For muon neutrino candidates, the backgrounds from NC neutrino interactions, $N^{\rm NC\,\nu}$, is estimated to be \NncContami. 
No NC contamination is found in the simulation, which is equivalent to the integrated luminosity of 600~$\mathrm{fb^{-1}}$ in the full FASER$\nu$ detector, for electron neutrino candidates. 
Therefore, the background assuming the fluctuations from the poisson probability to obtain 0 events in the simulation, normalized by the ratio of the expected and simulated NC events, is used. The NC background for electron neutrino candidates is estimated to be \BackgroundRandomNuENC.

To infer the neutrino flux, an energy-averaged signal strength $\mu$, 
common to both neutrinos and antineutrinos, is defined by $\phi_{\rm obs}^{{\nu}}= \mu\,\phi_{\rm predict}^{\nu}$,
where $\phi_{\rm predict}^{\nu}$ denotes the predicted flux for $\nu$. Accordingly, the total observed event count can be written as $N_{\rm exp}=\mu\,N_{\nu}^{\rm predict} + N^{\rm NH} + N^{\rm NC\,\nu},$
where $N_{\nu}^{\rm predict}$ denotes the predicted number of CC neutrino events in Eq.~\ref{eq:number_neutrino}. After determining $\mu$, the total number of neutrino interactions, $N^{\rm int}$, is computed for comparison with the flux predictions as
\begin{equation}
    N^{\rm int}
    \;=\;
    \mu\,
    \frac{L\,\rho\,l}{m_{\mathrm{nucleon}}}\,
    \int \mathrm{d}^2 x\,\mathrm{d}E 
    \bigl[\phi^{\nu}(E, \vec{x})\,\sigma^{\nu}(E, \vec{x})
    \;+\;\phi^{\bar{\nu}}(E, \vec{x})\,\sigma^{\bar{\nu}}(E, \vec{x})\bigr].
    \label{eq:interaction_rate}
\end{equation}

$\mu$ is estimated by constructing a likelihood function and sampling its posterior distribution via the Markov Chain Monte Carlo (MCMC) technique~\cite{gregory_2005}. 
The likelihood function for observing $N_{\rm obs}$ candidates is defined as
\begin{align}
  L =  P\bigl(N_{\rm obs} \,\big|\; N_{\rm exp}\bigr)\times 
  &\prod_{j} G_j \prod_{k}P_k,
  \label{eq:likelihood_number_nu_mu}
\end{align}
where $P(N_{\rm obs}|N_{\rm exp})$ is the Poisson probability of observing $N_{\rm obs}$ events given the expectation $N_{\rm exp}$.
The term $G_j$ denotes Gaussian priors corresponding to nuisance parameters from source $j$.
The nuisance parameters account for the following factors: the systematic uncertainty on neutral hadron production, the flux uncertainty from hadron production, and other systematic uncertainties in neutrino CC and NC predictions. 
The term $P_k$ denotes Poisson priors corresponding to nuisance parameters from statistical fluctuations of the background Monte Carlo (MC) predictions for the neutral hadron and NC contamination. 
Because the number of observed events is small, the Bayesian method is adopted to evaluate $\mu$. The posterior probability of $\mu$ is obtained by performing MCMC sampling using the Metropolis–Hastings algorithm~\cite{gregory_2005}.
The initial values and covariance matrix of the parameters are obtained from a maximum likelihood fit using MIGRAD implemented in the \texttt{iminuit} package~\cite{Dembinski2025-fp}. A total of $2.03\times10^5$ steps were generated, and the first 3000 steps were discarded as burn-in.

Since Eq.~\ref{eq:interaction_rate} integrates over energy, biases in $N^{\rm int}$ may arise from energy-dependent variations in both the flux and the detection efficiency. This effect was estimated by repeating the analysis using nine flux models.
Each of these models was made by combining light hadron predictions from EPOS-LHC, QGSJET II-04, SIBYLL 2.3d and predictions for charm production with POWHEG+Pythia using three choices of the factorization and renormalization scale~\cite{Buonocore:2023kna}. Although this bias was found to be almost negligible, this effect is included in the systematic uncertainty.
The number of neutrino interactions in the FASER$\nu$ detector is measured to be 
\[
  N^{\rm int}(\nu_e + \bar{\nu}_e) = 12.2^{+8.7}_{-6.4}  
  \qquad \text{and} \qquad
  N^{\rm int}(\nu_\mu+\bar{\nu}_\mu) = 36.0^{+16.1}_{-13.2} \ . 
\] 
The dominant sources of uncertainty correspond to the statistical uncertainty and the neutrino detection efficiency uncertainty for electron neutrinos and the neutrino detection efficiency uncertainty for muon neutrinos.

\section{Results}



\begin{figure}
\centering
\includegraphics[width=0.99\linewidth]{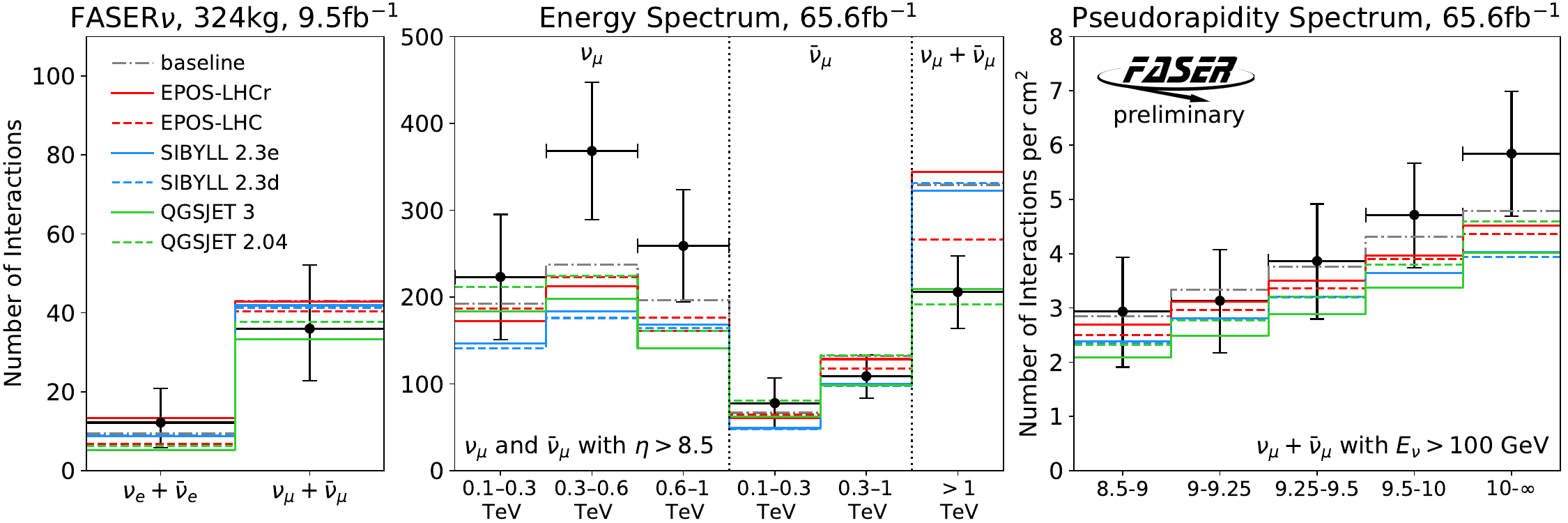}
\caption{Comparison of Cosmic Ray Monte Carlos to FASER data. The number of neutrino interactions in the respective target volumes as measured by FASER$\nu$ (left), as measured by FASER as a function of the neutrino energy (center), and by FASER as a function of rapidity (right) are shown. The measured values are compared to prediction by \texttt{EPOS-LHC}~\cite{Pierog:2013ria}, \texttt{EPOS-LHCr}~\cite{Pierog:2023ahq}, \texttt{SIBYLL~2.3d}~\cite{Riehn:2019jet}, \texttt{SIBYLL~2.3e}, \texttt{QGSJET~2.04}~\cite{Ostapchenko:2010vb}, \texttt{QGSJET~3}~\cite{Ostapchenko:2024myl}, and the FASER baseline model (see texts). The analyzed area on FASER$\nu$ corresponds to the pseudo-rapidity region of $\eta \geq 8.76$. 
The majority of the systematic uncertainties of FASER$\nu$ data points are correlated. Data points in the center and right panels are correlated because they use the same dataset.
}
\label{fig:models}
\end{figure}

The neutrino event rates measured by the FASER experiment, presented in Ref.~\cite{FASER:2024ref,FASER:2927629} and above, are compared with predictions made using various Monte Carlo tools. \cref{fig:models} compares measurements with predictions by \texttt{EPOS-LHC}, \texttt{EPOS-LHCr}, \texttt{SIBYLL~2.3d}, \texttt{SIBYLL~2.3e}, \texttt{QGSJET~2.04} and \texttt{QGSJET~3}. These predictions were obtained using the interface software packages \texttt{chromo}~\cite{Dembinski:2023esa} (for \texttt{QGSJET~3}) and \texttt{CRMC}~\cite{crmc201} (all the others). 
The baseline model of the FASER experiment, which uses \texttt{EPOS-LHC} for light hadrons and perturbative QCD-based \texttt{POWHEG+PYTHIA} for charm hadrons~\cite{FASER:2024ykc}, is also shown as a gray dash-dotted line. 
\texttt{EPOS-LHC}, \texttt{QGSJET~2.04} and \texttt{QGSJET~3} do not include a modeling for charm hadron production, which should be considered when comparing their predictions to data. 

\begin{figure}
\centering
\includegraphics[width=0.99\linewidth]{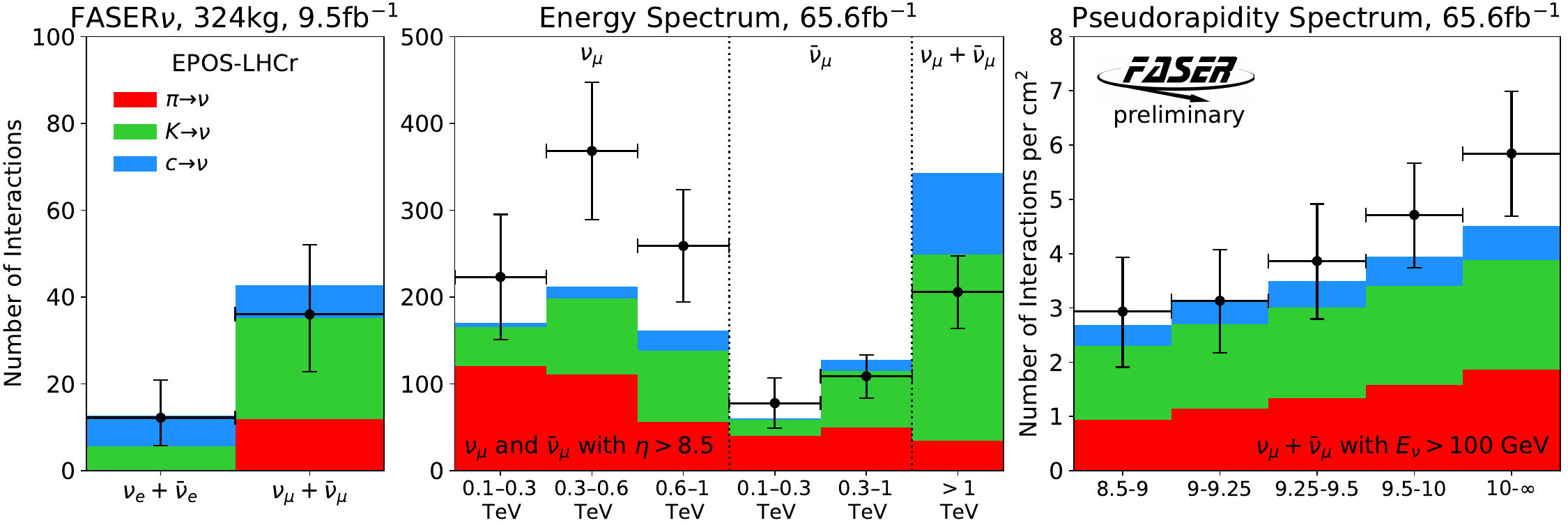}
\caption{Comparison of \texttt{EPOS-LHCr}, split by production modes, to FASER data. }
\label{fig:epos_r}
\end{figure}

The predictions are generally consistent. In all cases, there is an excess of neutrino events (but not anti-neutrino events) at energies between 300~GeV to 600~GeV in data compared to the generator predictions. In addition, generators including charm hadron production predict more high-energy neutrinos than observed. 
\texttt{EPOS-LHC}, \texttt{QGSJET~2.04}, and \texttt{QGSJET~3}, which -- unlike the other models considered -- do not include charm hadron production, show better agreement with the $\nu_\mu+\bar{\nu}_\mu$ prediction above 1 TeV.
\cref{fig:epos_r} shows the spectra predicted by \texttt{EPOS-LHCr}, the latest model of the \texttt{EPOS} series, split into neutrino production modes: pion decays, kaon decays, and charm decays. 
\texttt{EPOS-LHCr} overestimates the event rate of $\nu_\mu + \bar{\nu}_\mu$ above $1~\mathrm{TeV}$, where kaon and charm mesons contribute. However, it underestimates neutrino events at energies between 300~GeV and 600~GeV, where $\pi^+$ and kaons contribute. This indicates that the FASER results can further improve the latest hadronic interaction models.

\section{Conclusion and future prospects}


In this study, electron and muon neutrino event rates using the latest FASER$\nu$ results are interpreted, and measurements by the FASER experiment are compared with predictions from recent hadronic interaction models.
Discrepancies in a few energy bins of muon neutrinos are found. 
This motivates upcoming measurements by FASER, which will provide crucial information for understanding these discrepancies. 

Only a subset of collected data has been analyzed: 33\% of the electronic detector data and 3\% of the emulsion detector data collected up to the end of 2024.
More than $150~\mathrm{fb^{-1}}$ of proton-proton collision data is expected from 2025 to 2026, which will significantly increase the data set. 
With the inclusion of additional data, more precise flux measurements will be reported soon, enabling the validation of pion, kaon, and charm meson production with finer binning, reduced uncertainties, and multi-differential analyses.

FASER is approved to continue data collection in LHC-Run4 with an additional 680~$\mathrm{fb^{-1}}$ of data expected from 2030 to 2033. Various detector upgrades are under study for this, including larger electronic detectors on the collision axis and off axis ~\cite{FASER:2025myb}. 
R\&D projects for these proposed detectors are on-going.
Moreover, a new experimental area, the Forward Physics Facility, is proposed. Experiments with much larger target mass are proposed in the Forward Physics Facility, which will allow us to measure $O(10^5)$ electron neutrinos, $O(10^6)$ muon neutrinos, and $O(10^3)$ tau neutrinos~\cite{Adhikary:2024nlv,FPFICRC2025}. 
These measurements will uniquely constrain forward hadron production and ultimately shed light on the cosmic-ray muon puzzle. 
The large data set will also allow us to investigate parton distribution functions in regions of parameter space which other experiments cannot explore.


\bibliographystyle{JHEP}
\bibliography{references}

\clearpage
\section*{Full author list}


\noindent
\small
{
Roshan Mammen Abraham$^{1}$,
Xiaocong Ai$^{2}$,
Saul Alonso Monsalve$^{3}$,
John Anders$^{4}$,
Claire Antel$^{5}$,
Akitaka Ariga$^{6,7}$,
Tomoko Ariga$^{8}$,
Jeremy Atkinson$^{6}$,
Florian~U.~Bernlochner$^{9}$,
Tobias Boeckh$^{9}$
% ...
}

\begin{list}{}%
  {\setlength{\leftmargin}{1.5em}   
  \setlength{\itemsep}{1pt}      
  \setlength{\parsep}{0pt}       
  \setlength{\topsep}{3pt}       
  \setlength{\labelwidth}{1.5em}   
  \setlength{\labelsep}{0.5em}     
}
\small 
\item[$^{1}$] \textit{Department of Physics and Astronomy, University of California, Irvine, CA 92697-4575, USA}
\item[$^{2}$] \textit{School of Physics, Zhengzhou University, Zhengzhou 450001, China}
\item[$^{3}$] \textit{ETH Zurich, 8092 Zurich, Switzerland}
\item[$^{4}$] \textit{University of Liverpool, Liverpool L69 3BX, United Kingdom}
\item[$^{5}$] \textit{D\'epartement de Physique Nucl\'eaire et Corpusculaire, University of Geneva, CH-1211 Geneva 4, Switzerland}
\item[$^{6}$] \textit{Albert Einstein Center for Fundamental Physics, Laboratory for High Energy Physics, University of Bern, Sidlerstrasse 5, CH-3012 Bern, Switzerland}
\item[$^{7}$] \textit{Department of Physics, Chiba University, 1-33 Yayoi-cho Inage-ku, 263-8522 Chiba, Japan}
\item[$^{8}$] \textit{Kyushu University, 744 Motooka, Nishi-ku, 819-0395 Fukuoka, Japan}
\item[$^{9}$] \textit{Universit\"at Bonn, Regina-Pacis-Weg 3, D-53113 Bonn, Germany}
\item[$^{10}$] \textit{CERN, CH-1211 Geneva 23, Switzerland}
\item[$^{11}$] \textit{Nikhef National Institute for Subatomic Physics, Science Park 105, 1098 XG Amsterdam, Netherlands}
\item[$^{12}$] \textit{Institute for Particle Physics, ETH Z\"urich, Z\"urich 8093, Switzerland}
\item[$^{13}$] \textit{Department of Physics, Tsinghua University, Beijing, China}
\item[$^{14}$] \textit{INFN Sezione di Genova, Via Dodecaneso, 33--16146, Genova, Italy}
\item[$^{15}$] \textit{University of Oregon, Eugene, OR 97403, USA}
\item[$^{16}$] \textit{Affiliated with an international laboratory covered by a cooperation agreement with CERN.}
\item[$^{17}$] \textit{Institute of Space Science---INFLPR Subsidiary, Bucharest, Romania}
\item[$^{18}$] \textit{Department of Physics, University of Washington, PO Box 351560, Seattle, WA 98195-1460, USA}
\item[$^{19}$] \textit{Royal Holloway, University of London, Egham, TW20 0EX, United Kingdom}
\item[$^{20}$] \textit{Department of Physics and Astronomy, Technion---Israel Institute of Technology, Haifa 32000, Israel}
\item[$^{21}$] \textit{II.~Physikalisches Institut, Universität Göttingen, Göttingen, Germany}
\item[$^{22}$] \textit{Deutsches Elektronen-Synchrotron DESY, Notkestr.~85, 22607 Hamburg, Germany}
\item[$^{23}$] \textit{Department of Particle Physics and Astrophysics, Weizmann Institute of Science, Rehovot 76100, Israel}
\item[$^{24}$] \textit{Institut f\"ur Physik, Universität Mainz, Mainz, Germany}
\item[$^{25}$] \textit{Department of Physics \& Astronomy, University of Sussex, Sussex House, Falmer, Brighton, BN1 9RH, United Kingdom}
\item[$^{26}$] \textit{Nagoya University, Furo-cho, Chikusa-ku, Nagoya 464-8602, Japan}
\item[$^{27}$] \textit{University of Manchester, School of Physics and Astronomy, Schuster Building, Oxford Rd, Manchester M13 9PL, United Kingdom}
\item[$^{28}$] \textit{Dipartimento di Fisica ``Ettore Pancini'', Universit\`a di Napoli Federico II, Complesso Universitario di Monte S.~Angelo, I-80126 Napoli, Italy}
\item[$^{29}$] \textit{National Institute of Technology (KOSEN), Niihama College, 7-1, Yakumo-cho Niihama, 792-0805 Ehime, Japan}
\item[$^{30}$] \textit{Particle Physics Department, STFC Rutherford Appleton Laboratory, Harwell Campus, 
Didcot, OX11 0QX, United Kingdom}
\end{list}

\end{document}